\documentclass[journal]{IEEEtran}
\usepackage{graphicx}
\usepackage{amsfonts}
\usepackage{mathrsfs}
\usepackage{amssymb,amsmath}
\usepackage{bm}
\usepackage{algorithm}
\usepackage{algorithmic}
\usepackage{theorem}
\usepackage{array,color}
\usepackage{cite}
\usepackage{stfloats}
\usepackage{subfigure}
\usepackage{array}
\usepackage{multirow}
\usepackage{bbding}
\usepackage{epstopdf}
\newcommand{\tb}{\mathbf}

\begin{document}
 	\title{{Location-aware Predictive Beamforming for UAV Communications: A Deep Learning Approach}}
 		\author{{Chang Liu, \emph{Member, IEEE}, Weijie Yuan, \emph{Member, IEEE}, Zhiqiang Wei, \emph{Member, IEEE}, Xuemeng Liu, and Derrick Wing Kwan Ng, \emph{Senior Member, IEEE}}

 \thanks{The authors are with the School of Electrical Engineering and Telecommunications, University of New South Wales, NSW 2052, Australia.}



 }
\maketitle

\begin{abstract}
Unmanned aerial vehicle (UAV)-assisted communication becomes a promising technique to realize the beyond fifth generation (5G) wireless networks, due to the high mobility and maneuverability of UAVs which can adapt to heterogeneous requirements of different applications.
However, the movement of UAVs impose challenge for accurate beam alignment between the UAV and the ground user equipment (UE).
In this letter, we propose a deep learning-based location-aware predictive beamforming scheme to track the beam for UAV communications in a dynamic scenario.
Specifically, a long short-term memory (LSTM)-based recurrent neural network (LRNet) is designed for UAV location prediction. Based on the predicted location, a predicted angle between the UAV and the UE can be determined for effective and fast beam alignment in the next time slot, which enables reliable communications between the UAV and the UE.
Simulation results demonstrate that the proposed scheme can achieve a satisfactory UAV-to-UE communication rate, which is close to the upper bound of communication rate obtained by the perfect genie-aided alignment scheme.
\end{abstract}
\section{Introduction}
Unmanned aerial vehicle (UAV)-assisted communication, where a UAV is deployed as a base station (BS) or a relay has become one of the key enabling techniques for enhancing the network coverage, energy-efficiency, and capacity in the beyond fifth generation (5G) wireless networks \cite{wong2017key}.
Relying on its high maneuverability, a UAV can support users in different areas to fulfill various applications, such as photography, mapping, and surveying \cite{xu2020multiuser}.
In most practical scenarios, multi-antenna systems are adopted to facilitate UAV communications.
Therefore, the beamforming technique, through which an antenna array can establish directional energy-focusing beams at specific angles to improve the communication efficiency, is of great importance for high data-rate transmission in UAV-assisted communications and thus it has attracted extensive attention from both academia and industry \cite{zeng2019accessing}.

In the literature, effective algorithms and schemes have been proposed for addressing the beamforming problem in UAV-assisted communications.
For instance, in \cite{zhao2018beam}, a hybrid beam tracking approach was proposed for the beam alignment in UAV-to-satellite links. In the proposed scheme, several senors are mounted on the UAV to obtain navigation information for beam alignment, which however, are not always practical due to the size and power limitation on UAVs.
In addition, \cite{miao2020lightweight} proposed a position-based lightweight beamforming scheme for 5G UAV broadcasting communications, where a global positioning system (GPS)-assisted beam tracking approach was studied for the beam alignment in BS-to-UAVs links. Nevertheless, the deployment of the GPS is costly and complicated, which is not always realizable for UAVs.
As an alternative, pilot-based methods have been developed to facilitate beam alignment. For example, the authors in \cite{yang2019beam} jointly adopted both the beam training and the velocity estimation to optimize a pilot-based beam tracking scheme for beam alignment.
Also in \cite{huang20203d}, a three-dimensional beamforming algorithm was designed by applying the dynamic pilot insertion.
However, for high-speed UAVs, due to the relatively long delay required in pilot transmission, these pilot-based methods cannot capture the up-to-date UAV locations in a timely manner and only achieve a limited beam tracking performance.
More importantly, the high maneuverability of UAVs introduces a wide possible range of trajectory, which complicates the task of beam alignment. Hence, a more practical beam alignment scheme, which can response timely and is suitable for various UAV trajectories, is desired.

In practice, the trajectory of a UAV can be treated as a temporal correlated sequence, while the deep learning (DL) technique \cite{goodfellow2016deep}, especially the long short-term memory (LSTM) neural network, has powerful capability in exploiting the temporal dependencies.
Motivated by this, in this letter, we propose a DL-based location-aware predictive beamforming scheme to facilitate the beam alignment in UAV-assisted communications.
In the proposed scheme, a LSTM-based recurrent neural network (LRNet) is specifically designed to exploit the temporal dependency of the UAV trajectory sequences for UAV location prediction.
In contrast to existing beam alignment methods, e.g., \cite{zhao2018beam, miao2020lightweight, yang2019beam, huang20203d}, the proposed scheme neither requires the use of pilots nor navigation sensors. Instead, only a LRNet-based offline training is required to exploit the temporal features of the UAV trajectory for accurate location prediction \cite{hochreiter1997long}.
Based on the predicted location, a predicted angle between the UAV and the user equipment (UE) can be determined for the beam alignment in the next time slot, which enables reliable UAV communications.
Simulation results demonstrate that the proposed predictive beamforming scheme can accurately predict the UAV location for beam alignment and achieve a stable communication rate for the UAV-to-UE link.

	\emph{Notations:} Terms $\mathbb{R}$ and $\mathbb{C}$ are the sets of the real numbers and the complex numbers, respectively. The superscripts $T$ and $H$ are used to represent the transpose and the conjugate transpose, respectively. $|\cdot|$ and $||\cdot||$ denote the modulus of a complex number and the Euclidean norm of a vector. $\mathbf{I}$ and $\mathbf{0}$ denote the identity matrix and the zero vector, respectively. $U(a,b)$ denotes the uniform distribution with the range between $a$ and $b$ ($a\leq b$). $\mathcal{N}(\bm{\mu},\mathbf{\Sigma})$ denotes the Gaussian distribution with mean vector $\bm{\mu}$ and covariance matrix $\mathbf{\Sigma}$. $\mathrm{arccos}(\cdot)$ is the inverse cosine function.

\section{System Model}
\begin{figure}[t]
\centering
\includegraphics[width=.4\textwidth]{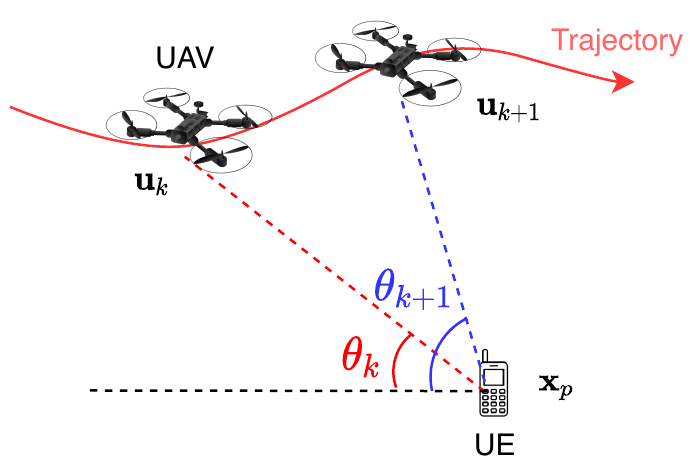}\vspace{-0.3cm}

\caption{{UAV communication model.}}%
\label{UAV_position}%
\vspace{-0.6cm}
\end{figure}

The considered UAV-to-UE communication scenario is illustrated in Fig. \ref{UAV_position}. The UAV is equipped with an $M$-antenna uniform linear array (ULA) hovering in the sky and communicating with a static UE with an $N$-antenna ULA. For ease of exposition, we assume that the ULAs of the UAV and the UE are in parallel, then the angle-of-arrival (AOA) is identical to the angle-of-departure (AoD) \cite{zhao2018beam}. Without loss of generality, the coordinate of the UE is set as $\tb{x}_p=[x_p,y_p]^{\rm T}$, where $x_p$ and $y_p$ are the coordinates on $x$-axis and $y$-axis, respectively. The location of the moving UAV at time instant $k$ is denoted by $\mathbf{u}_k = [x_k,y_k]^{\rm T}$. Consequently, the angle of the UAV relative to the UE at time $k$ is given by
\begin{align}\label{theta_k}
\theta_k = \arccos \frac{x_k-x_p}{\|\mathbf{u}_k-\tb{x}_p\|},
\end{align}
as shown in Fig. \ref{UAV_position}.

For practical UAV communications, a rich scattering scenario rarely appears \cite{xu2020multiuser}, therefore, the UAV-to-UE channel consists of only the line-of-sight (LOS) communication link, denoted by a channel matrix $\tb{H}_k\in \mathbb{C}^{N\times M}$, i.e.,
\begin{align}
\tb{H}_k= \frac{c}{4\pi f_c \|\mathbf{u}_k-\tb{x}_p\|}\tb{b}(\theta_k)\tb{a}(\theta_k)^{\rm H}
\end{align}
where $c$ is the signal propagation speed, $f_c$ is the centre carrier frequency, and $\tb{a}(\theta_k)$ and $\tb{b}(\theta_k)$ denote the normalized transmitting and receiving steering vectors, respectively. In particular, the antenna spacing for ULAs of both the UAV and the UE can be set as $d_c = \frac{f_c}{2c}$. Therefore, we have
\begin{align}\label{tx_steer}
\tb{a}(\theta_k) &= \sqrt{\frac{1}{M}}\left[1, e^{-j{\frac{2\pi d_c \cos\theta_k }{\lambda_c} }},...,e^{-j{\frac{2\pi d_c (M-1) \cos\theta_k }{\lambda_c}}}\right]^{\rm T},\\\label{rx_steer}
\tb{b}(\theta_k) &= \sqrt{\frac{1}{N}}\left[1, e^{-j{\frac{2\pi d_c \cos\theta_k }{\lambda_c}}},...,e^{-j{\frac{2\pi d_c (N-1) \cos\theta_k }{\lambda_c}}}\right]^{\rm T},
\end{align}
where we set $\lambda_c = 2d_c$ for simplifying the steering vectors.

Assume that $k$ denotes the current time slot, let us denote the transmitted signal from the UAV-mounted BS at time $k$ by $s_k \in\mathbb{C}$. The transmitted signal is steered towards the intended direction via a transmit beamformer $\tb{f}_k$ and propagates through the channel $\tb{H}_k$. After receiving the signal, the UE adopts a receive beamformer $\tb{w}_k$ and obtains the received sample $r_{k}$ as
\begin{align}\label{receive_signal}
{r}_{k}= \tb{w}_k^{\rm H} \tb{H}_k \tb{f}_k s_{k} + \eta_{k},
\end{align}
where $\eta_{k}\in\mathbb{C}$ is the noise variable with zero mean and variance $\sigma^2$.
In general, the transmit beamformer adopted at the UAV can be designed based on the relative angle $\theta_k$, i.e., $\tb{f}_k = \tb{a}({\theta}_k)$ since the UAV can obtain its location information at each time slot.
In contrast, the UE can only obtain an estimated angle $\tilde{\theta}_{k}$ based on the previous UAV locations, i.e., $\{{\tb{u}}_{k-L},...,{\tb{u}}_{k-1}\}$ which are sent by the UAV at the previous time slots. Based on $\tilde{\theta}_{k}$, the receive beamformer at the UE can be expressed as $\tb{w}_k = \tb{b}(\tilde{\theta}_k)$.
According to \eqref{receive_signal}, the receive signal-to-noise ratio (SNR) is given by
\begin{align}\label{SNR}
\textrm{SNR}_k &= \frac{p_t \left|h_k\tb{w}_k^{\rm H}\tb{b}(\theta)\tb{a}(\theta)^{\rm H}\tb{f}_k\right|^2}{\sigma^2}\nonumber\\
&= \frac{p_t \left|h_k\tb{w}_k^{\rm H}\tb{b}(\theta)\right|^2}{\sigma^2},
\end{align}
where $p_t = |s_{k}|^2$ denotes the transmit signal power and $h_k = \frac{c}{4\pi f_c\|\mathbf{u}_k-\tb{x}_p\|} $ is the path-loss gain at time $k$. The achievable rate of the UAV-to-UE link at time $k$ is therefore given by
\begin{align}
R_k = \log_2 \left(1+\textrm{SNR}_k\right).
\end{align}
We can observe from \eqref{SNR} that if the predicted angle $\tilde{\theta}_k$ equals to the actual angle $\theta_k$, the SNR as well as the achievable rate is maximized. However, due to the dynamic nature of the UAV, the angle $\theta_k$ is time-varying which imposes a challenge on obtaining accurate beam alignment. In the next section, we will propose a learning-based predictive beamforming algorithm to predict the UAV location at the UE side and consequently obtain the predicted angle $\tilde{\theta}_k$. Thus, the UAV and the UE can directly establish the communication link though beam alignment at time $k$.

\section{Location-aware Predictive Beamforming}
The proposed predictive beamforming scheme consists of the location prediction step and the beamforming design step. First, the UAV location is predicted at the UE via a neural network in the prediction step. Then, this information is adopted for beam alignment in the beamforming design step.

Generally, the movement of the UAV can be modeled as \cite{zeng2019accessing}
\begin{align}\label{UAV_mov}
\tb{u}_k = \tb{u}_{k-1} +\mathbf{v}_{k-1} \Delta T + \mathbf{\Lambda}_{k-1}.
\end{align}
Here, $\mathbf{v}_{k-1} = [v^{x}_{k-1}, v^{y}_{k-1}]^T, \forall k$, denotes the UAV average velocity at time $k-1$, where $v^{x}_{k-1}$ and $v^{y}_{k-1}$ are the projections on the $x$-axis and $y$-axis, respectively. Let $\alpha_{k-1}=|\mathbf{v}_{k-1}|$ and $\beta_{k-1}=\mathrm{arccos}(\frac{v^{x}_{k-1}}{\mathbf{v}_{k-1}})$ denote the amplitude and phase of $\mathbf{v}_{k-1}$, respectively.
Without loss of generality, the amplitude and the phase are assumed to follow the uniform distribution \cite{zeng2019accessing}, i.e., $\alpha_{k-1}\sim U(a,b)$ and $\beta_{k-1}\sim U(c,d)$, where $a,b,c,d$ are constants with $0 \leq a \leq b$ and $-\pi \leq c \leq d \leq \pi$.
Also, $\Delta T$ represents the time duration of a time slot.
In addition, considering that random wind gust may cause the UAV location deviation \cite{xu2020multiuser, yuan2020learning}, we then introduce $\mathbf{\Lambda}_{k-1}=[\lambda^x_{k-1},\lambda^y_{k-1}]^T$ to characterize the environment uncertainty at time $k-1$, where $\lambda^{x}_{k-1}$ and $\lambda^{y}_{k-1}$ are the uncertainty values on the $x$-axis and $y$-axis, respectively.
In addition, $\mathbf{\Lambda}_{k-1}\sim \mathcal{N}(\mathbf{0}, \sigma_v^2\mathbf{I})$ is assumed to be a Gaussian random vector and $\sigma_v^2$ is the variance depending on the wind \cite{xu2020multiuser}.

Now, we propose a DL approach for predicting the UAV location at time $k$, which relies on the previous $L$ locations, i.e., the location sequence $\{{\tb{u}}_{k-L},...,{\tb{u}}_{k-1}\}$ sent by the UAV.

\begin{figure}[t]
\centering
\includegraphics[width=\linewidth]{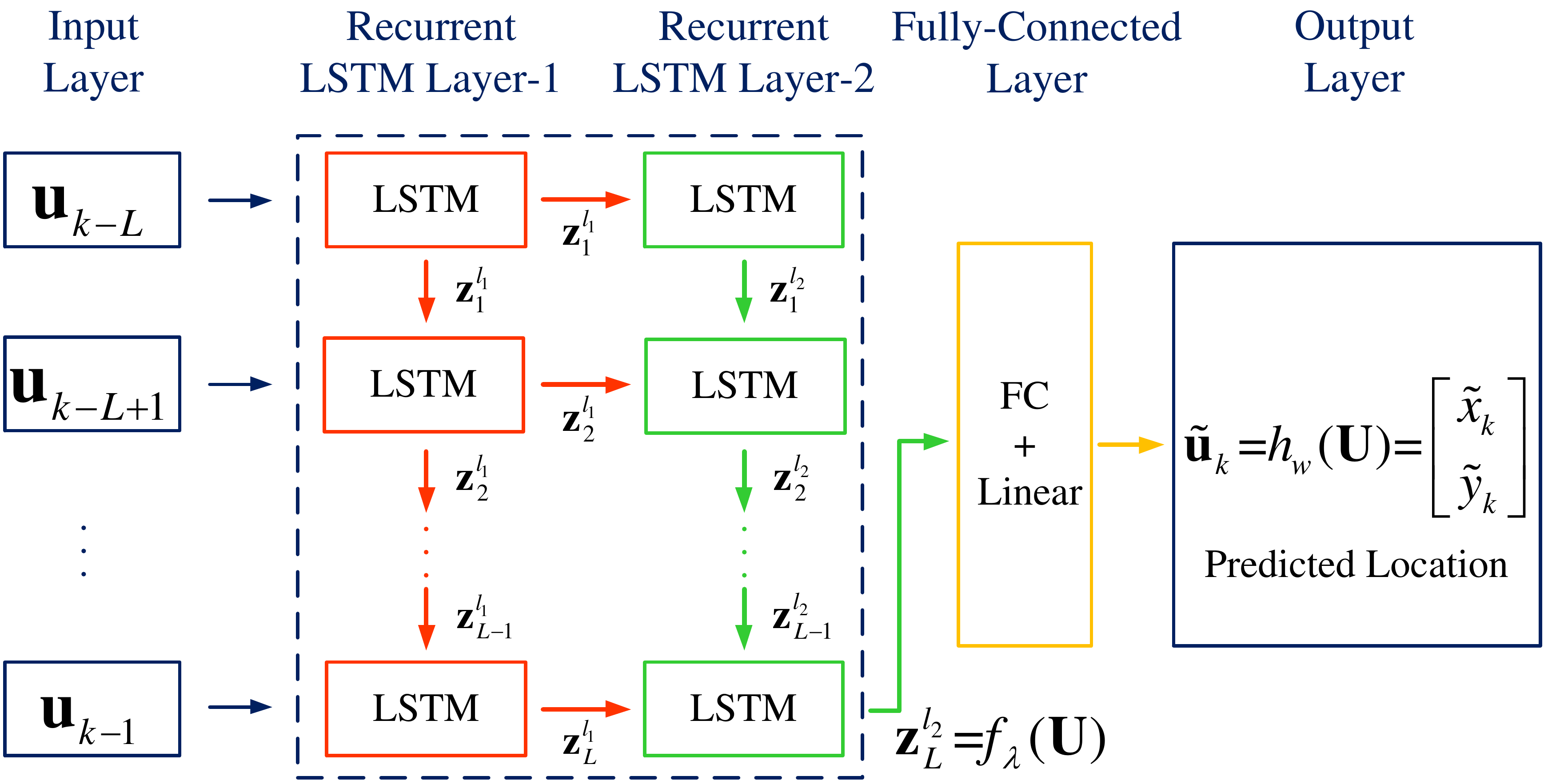}
\caption{{The proposed LRNet for location prediction. }}%
\label{Fig_LRNet}%
\end{figure}

\subsection{Deep Learning-based Location Prediction}
To fully exploit the temporal features of the UAV location sequence, a LRNet is proposed for location prediction.
As shown in Fig. \ref{Fig_LRNet}, the LRNet consists of an input layer, recurrent LSTM layer-1, recurrent LSTM layer-2, a fully-connected (FC) layer, and an output layer. The hyperparameters of the proposed LRNet are summarized in Table I and each layer will be introduced in the following.

\begin{table}[t]
\normalsize
\caption{Hyperparameters of the proposed LRNet}
\centering
\small
\renewcommand{\arraystretch}{1.2}
\begin{tabular}{c c c}
  \hline
   \multicolumn{3}{l}{\textbf{Input}: $\mathbf{U}$ with the size of $2 \times L$}  \\
  \hline
  \hspace{0.6cm} \textbf{Layers} & \textbf{Parameters} &  \hspace{0.6cm} \textbf{Values}   \\
  \hspace{0.6cm} LSTM layer-1 & Size of $\mathbf{z}_i^{l_1}$  & \hspace{0.6cm}  $ 50 \times 1 $   \\
  \hspace{0.6cm} LSTM layer-2  & Size of $\mathbf{z}_i^{l_2}$  & \hspace{0.6cm} $ 100 \times 1 $   \\
  \hspace{0.6cm} FC layer & Activatioin function & \hspace{0.6cm} Linear \\
  \hline
   \multicolumn{3}{l}{\textbf{Output}: $\tilde{\mathbf{u}}_k$ with the size of $2 \times 1$}  \\
  \hline
\end{tabular}
  \vspace{-0.6cm} \\
\end{table}

\emph{a) Input Layer:}
Denote by
\begin{equation}\label{}
  \mathbf{U}_k = [ {\tb{u}}_{k-L},...,{\tb{u}}_{k-1} ]
\end{equation}
the $L$-length location sequence. $\mathbf{U}_k \in \mathbb{R}^{2\times L}$ carries the temporal location information and thus is used as the input of the LRNet.

\emph{b) Recurrent LSTM Layers:}
To exploit the long-term dependencies of the location sequence, two recurrent LSTM layers are adopted, named recurrent LSTM layer-1 and recurrent LSTM layer-2, as shown in Fig. \ref{Fig_LRNet}.
For each recurrent layer, there are $L$ identical LSTM modules cascaded handling the input from the input layer in the past $L$ time steps, respectively. Besides, let $\mathbf{z}_i^{l_j}$ represent the output of the LSTM of the $j$-th, $j\in\{1,2\}$, layer at the $i$-th $i\in\{1,2,\cdots,L\}$ time step, we can then characterize the LSTM modules of different layers.
For LSTM layer-1, the output $\mathbf{z}_i^{l_1}$ is not only sent to the next recurrent layer, but also saved as the input of the LSTM module at the next time step.
For LSTM layer-2, the output $\mathbf{z}_i^{l_2}$ is only served as the input of the same LSTM module at the next time step. In particular, the $L$-th output $\mathbf{z}_L^{l_2}$ is the output of the LSTM layers which is expressed as
\begin{equation}\label{}
  \mathbf{z}_L^{l_2} = f_{\lambda}(\mathbf{U}_k),
\end{equation}
where $f_{\lambda}(\cdot)$ with parameters $\lambda$ denotes a non-linear function characterizing the operation from LSTM layer-1 to LSTM layer-2.

\emph{c) Fully-connected Layer:}
After the feature extraction from the two recurrent LSTM layers, a FC layer is then added between the recurrent LSTM layer-2 and the output layer to linearly combine these extracted features to further improve the network performance \cite{liu2019deep}.

\emph{d) Output Layer:}
Based on the above analysis, the output of the LRNet can be expressed as
\begin{equation}\label{u_k_LRNet}
  \tilde{\mathbf{u}}_k = \sigma(\mathbf{W}f_{\lambda}(\mathbf{U}_k)+\mathbf{b}),
\end{equation}
where $\tilde{\mathbf{u}}_k \in \mathbb{R}^{2 \times 1}$ is the predicted location at time $k$ by LRNet, $\mathbf{W}\in \mathbb{R}^{2 \times 100}$ and $\mathbf{b} \in \mathbb{R}^{2 \times 1}$ are the weights and bias of the FC layer, and $\sigma(\cdot)$ represents the linear activation function.
For simplicity, we can rewrite (\ref{u_k_LRNet}) as
\begin{equation}\label{}
  \tilde{\mathbf{u}}_k = h_w(\mathbf{U}_k) =
  \left[
  {\begin{array}{*{20}{c}}
  {\tilde{x}_k}\\
  {\tilde{y}_k}
  \end{array}}
  \right],
\end{equation}
where $h_w(\cdot)$ is the total expression of LRNet with parameters $w=\{\lambda,\mathbf{W},\mathbf{b}\}$ and $\tilde{x}_k$ and $\tilde{y}_k$ are the elements of the predicted location as defined in (\ref{theta_k}).

In fact, the location prediction of (\ref{UAV_mov}) is essentially a regression problem. Inspired by this, a mean square error (MSE) cost function is selected for the proposed LRNet, which is expressed as \cite{goodfellow2016deep}
\begin{equation}\label{mse_cost_function}
  J_{\rm MSE}(w) = \frac{1}{2N}\sum\limits_{n=1}^{N} \left(\mathbf{u}_k^{(n)} - h_w(\mathbf{U}_k^{(n)})  \right)^2,
\end{equation}
where $\mathbf{u}_k^{(n)}$ and $\mathbf{U}_k^{(n)}$ represent the label and the input of the $n$-th, $n\in\{1,2,\cdots,N\}$, training example.
Therefore, we can then adopt a backpropagation algorithm \cite{goodfellow2016deep} to update the parameters progressively to minimize $J_{\mathrm{MSE}}$ in (\ref{mse_cost_function}) and finally obtain the well-trained LRNet for location prediction.

\subsection{Predictive Beamforming Design}
Based on the predicted location $\tilde{\mathbf{u}}_k$ from the LRNet, the UE is able to determine the predicted angle $\tilde{\theta}_k$ as
\begin{align}\label{theta_k_w}
\tilde{\theta}_k = \arccos \frac{\tilde{x}_k-x_p}{\|\tilde{\mathbf{u}}_k-\tb{x}_p\|},
\end{align}
and formulate the receive beam as $\tb{w}_k = \tb{b}(\tilde{\theta}_k)$. Based on this prediction mechanism, the UAV and the UE can directly establish the communication link at time $k$.

Moreover, the powerful DL-based prediction enables the UE to combat unexpected link failures. For example, given the previous states of the UAV, denoted by $\{\tb{u}_{k-L},...,\tb{u}_{k-1}\}$, the UE can not only predict the UAV location at time $k$, but also a further step to $\tilde{\tb{u}}_{k+1}$. In a special circumstance that the UAV fails to report its location to the UE at time $k$, the UE can still exploit the predicted location $\tilde{\tb{u}}_{k+1}$ to formulate the receive beamforming at time $k+1$ for maintaining a reliable UAV-to-UE communication.

\section{Simulation Results}
This section presents the simulation results to verify the effectiveness of the proposed scheme. For brevity, we assume that the location of the UE is fixed at $\tb{x}_p = [0,0]^{\rm T}$. The ULA mounted on the UAV has $M=16$ antennas while the UE is equipped with $N=8$ antennas. The operating frequency for the considered UAV-UE system is $f_c = 30$ GHz. The transmit power $p_t$ is set as $20$ dBm and the thermal noise power is $\sigma^2 = -90$ dBm. We consider a total number of $200$ time instants and the duration for each time instant is $\Delta T = 0.02$ s. The UAV movement model is given in (\ref{UAV_mov}) with $\sigma_v=0.01$, and we set $a= 0.4~\mathrm{m}/\Delta T$, $b= 0.7~\mathrm{m}/\Delta T$, $c = -\pi/6$, and $d = \pi/6$, i.e., the UAV has an equivalent speed around $20 \sim 35~\mathrm{m/s}$. Finally, we assume that the input length $L=20$.

\begin{figure}[t]
\centering
\includegraphics[width=0.4\textwidth]{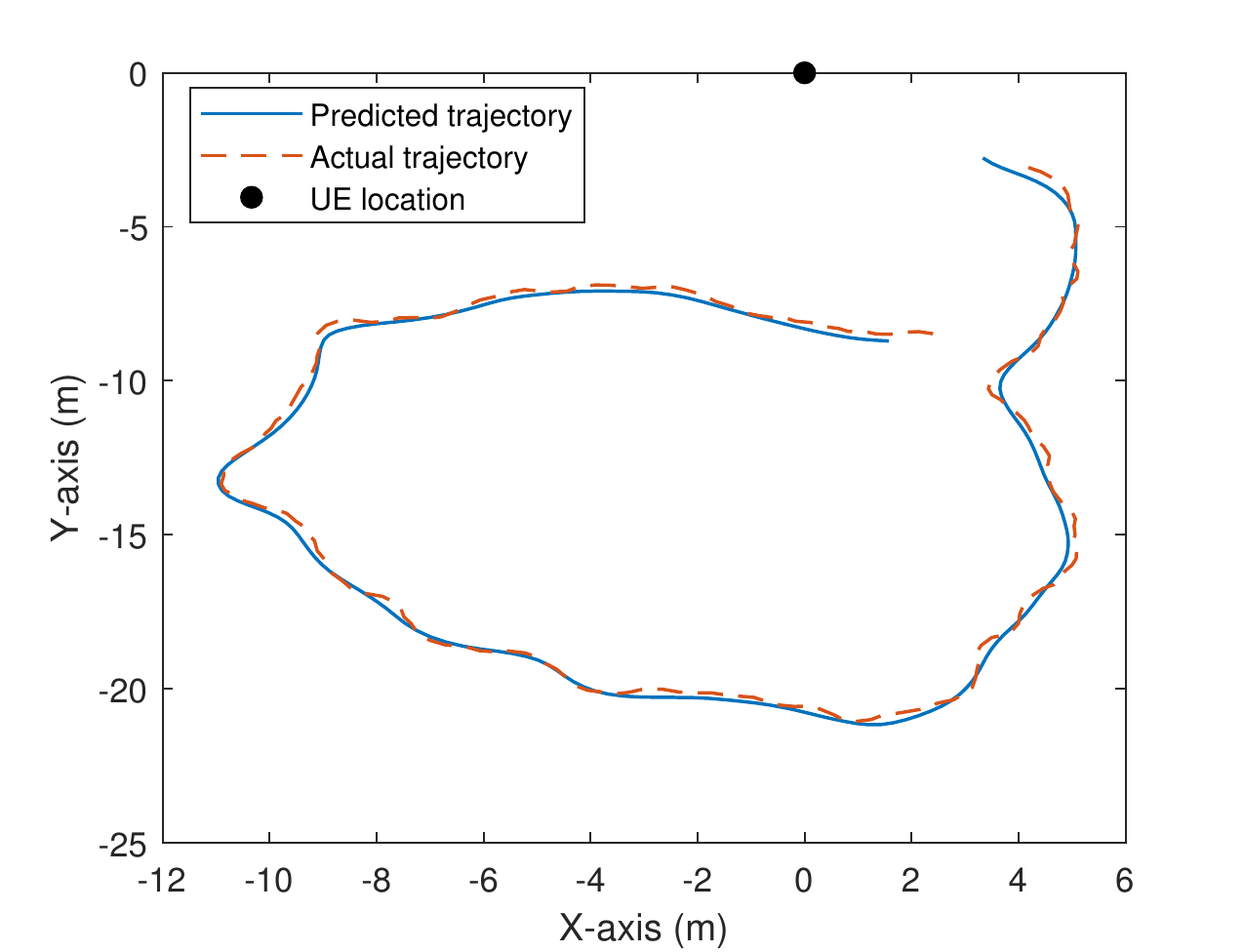}
\vspace{-0.15cm}
\caption{{Illustration of the actual and the predicted trajectories. }}%
\label{trajectory}
\vspace{-0.4cm}
\end{figure}

In Fig. \ref{trajectory}, we show the actual trajectory of the UAV and the trajectory determined by the predicted locations of the UAV. We can see that relying on the LRNet, the UE is capable of efficiently predicting the location of the UAV in real-time. Even though the moving direction of the UAV varies rapidly at some time instants, we can observe that the location prediction error is around the level of $0.1$ m at all time instants.

\begin{figure}[t]
\centering
\includegraphics[width=0.4\textwidth]{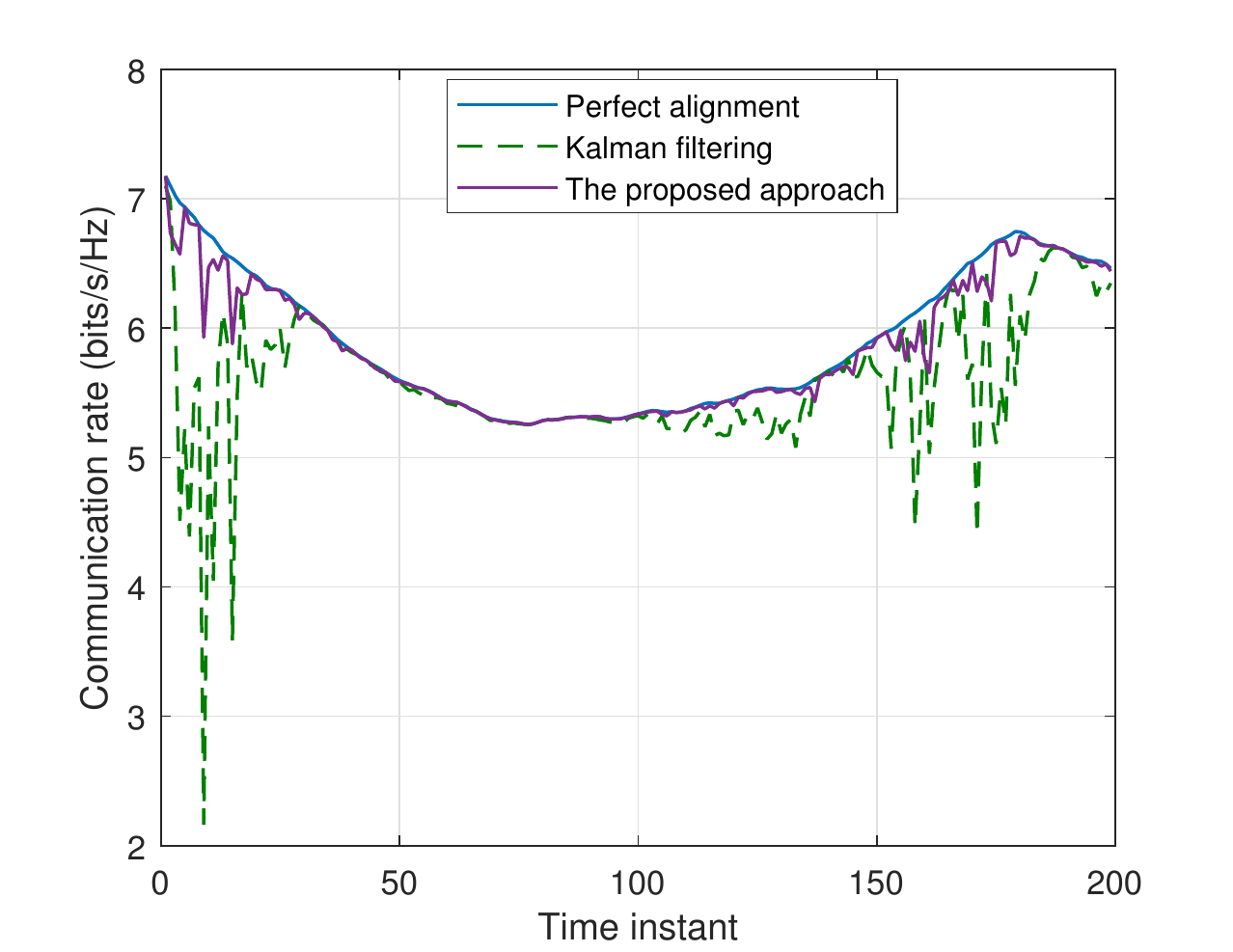}
\vspace{-0.15cm}
\caption{{The comparison of the communication rate between the baseline schemes and the proposed approach.}}%
\label{rate}
\vspace{-0.4cm}
\end{figure}

Fig. \ref{rate} depicts the communication rate of the proposed predictive beamforming approach. For comparison, we also plot the communication rates for two other schemes. Specifically, the genie-aided perfect alignment scheme assumes that the estimated angle $\tilde{\theta}_k$ is identical to the actual angle $\theta_k$ at each time instant. Obviously, this serves as the upper bound of the communication rate. Another baseline scheme predicts the UAV location via Kalman filtering based on the locations of the previous two time instants for determining the beamformer. We can observe from Fig. \ref{rate} that the proposed predictive beamforming scheme can establish a reliable communication link between the UAV and the UE. In particular, it can be seen that the rate of the proposed scheme decreases slightly at some instants due to a drastic change of the UAV location, which can be in general addressed by adopting a higher value of $L$ in the LRNet \cite{goodfellow2016deep}. On the contrary, although the Kalman filtering-based scheme has a very simple process for predicting the UAV location, it suffers from severe rate degradation and huge rate fluctuation due to the mismatch of the receive beamformer and the channel. Through this comparison, we can verify the effectiveness of the proposed approach in guaranteeing stable communications.

\section{Conclusions}
This letter studied the practical beam alignment problem for UAV-assisted communications and adopted a DL approach to develop a location-aware predictive beamforming scheme to track the beam for UAV communications in a dynamic scenario.
The proposed predictive beamforming scheme design consists of the DL-based location prediction and the predictive beamforming.
Specifically, a LRNet was designed to exploit the temporal features of the UAV trajectory for location prediction.
Based on the predicted location, a beam alignment was designed for the next time slot to enable reliable communications between the UAV and the UE.
Simulation results showed that the achievable communication rate of the proposed method approaches that of the upper bound of the communication rate obtained by the perfect alignment scheme.

\bibliographystyle{ieeetr}

\setlength{\baselineskip}{10pt}

\bibliography{ReferenceSCI2}

\end{document}